\documentclass[sigconf]{acmart}

\renewcommand\footnotetextcopyrightpermission[1]{}
\usepackage{bm}
\usepackage{tabularx}
\usepackage{booktabs}
\usepackage{natbib}
\usepackage[subrefformat=parens]{subcaption}
\theoremstyle{definition}

\newcommand{\pre}{\mathrm{pre}}
\newcommand{\ctx}{\mathrm{ctx}}
\newcommand{\nud}{\mathrm{nud}}

\newcommand{\dd}{\mathrm{d}}
\newcommand{\Nin}{\mathcal{N}^{\mathrm{in}}}
\newcommand{\Nout}{\mathcal{N}^{\mathrm{out}}}
\newcommand{\diag}{\mathrm{diag}}

\newcommand{\etal}{\textit{et al.,}}

\usepackage[linesnumbered,ruled,vlined]{algorithm2e}
\AtBeginDocument{%
  \providecommand\BibTeX{{%
    \normalfont B\kern-0.5em{\scshape i\kern-0.25em b}\kern-0.8em\TeX}}}

\begin{document}

\title{The Impact of Micro-level User Interventions on Macro-level Misinformation Spread}
\author{Satoshi Furutani}
\email{satoshi.furutani@ntt.com}
\orcid{1234-5678-9012}
\affiliation{%
  \institution{Social Informatics Laboratories, NTT,~Inc.}
  \streetaddress{3-9-11, Midori-cho, Musashino-shi}
  \city{Tokyo}
  \country{Japan}
  \postcode{180-8585}
}

\author{Toshiki Shibahara}
\affiliation{%
  \institution{Social Informatics Laboratories, NTT,~Inc.}
  \streetaddress{3-9-11, Midori-cho, Musashino-shi}
  \city{Tokyo}
  \country{Japan}
  \email{toshiki.shibahara@ntt.com}
}

\author{Mitsuaki Akiyama}
\affiliation{%
  \institution{Social Informatics Laboratories, NTT,~Inc.}
  \streetaddress{3-9-11, Midori-cho, Musashino-shi}
  \city{Tokyo}
  \country{Japan}
  \email{mitsuaki.akiyama@ntt.com}
}


\renewcommand{\shortauthors}{Furutani \etal}

\begin{abstract}
User interventions such as nudges, prebunking, and contextualization have been widely studied as countermeasures against misinformation, and shown to suppress individual users' sharing behavior.
However, it remains unclear whether and to what extent such individual-level effects translate into reductions in collective misinformation prevalence.
In this study, we incorporate user interventions as reductions in users' susceptibility within an empirically calibrated network-based misinformation diffusion model. 
We then systematically evaluate how intervention strength, scale, timing, target selection, and combinations of interventions affect overall misinformation prevalence through numerical simulations.
The simulation results reveal that current user-level interventions may not necessarily produce sufficient collective effects. 
Specifically, each intervention alone only modestly suppresses misinformation prevalence, and even when design adjustments such as expanding intervention scale, implementing interventions earlier, or strategically selecting target users are introduced, the resulting gains in suppression remain limited. 
Although combining multiple interventions improves the suppression effect compared to using each intervention alone, achieving substantial reductions in misinformation prevalence remains difficult within realistically attainable intervention levels.
This study quantitatively clarifies the gap between micro-level user interventions and macro-level misinformation spread, and demonstrates the limitations of evaluating misinformation countermeasures based solely on individual-level effectiveness.
\end{abstract}

\maketitle
\pagestyle{plain}

\section{Introduction}

The spread of misinformation on online social media platforms is recognized as a serious challenge that can have severe influence on political decision-making and public health. 
To suppress the spread of such misinformation, the research community have developed a variety of user interventions, including nudges, prebunking, warning labels, and debunking, as surveyed in~\cite{hartwig2024landscape,roozenbeek2023countering,kozyreva2024toolbox}. 
In addition, major platforms have introduced forms of \textit{contextualization} that provide supplementary context for potentially misleading content, including warning labels, fact-check notices, and crowdsourced annotations (e.g., Community Notes on Twitter/X). 
Numerous studies in cognitive science, human-computer interaction (HCI), and computational social science have shown that these interventions can influence individual-level truth discernment, sharing willingness, and engagement, based on user surveys and social media data analysis~\cite{basol2021towards,katsaros2022reconsidering,epstein2022explanations,jose2025s}.

However, the effectiveness of user intervention at the individual level does not necessarily guarantee its effectiveness  in suppressing misinformation prevalence at the collective level.
For example, even if an intervention reduces individuals' susceptibility to misinformation for a large fraction of users, diffusion may remain largely unaffected if the resulting change is insufficient to disrupt the underlying network diffusion dynamics.
Nevertheless, it remains insufficiently understood how individual-level intervention effects translate into collective-level misinformation prevalence through network diffusion.
In particular, it is unclear how the collective impact of interventions depends on design factors such as intervention strength, scale, timing, and the selection of target users.

Against this background, this study quantitatively evaluates how micro-level user interventions influence macro-level misinformation prevalence via diffusion on a network through numerical simulations.
We model misinformation diffusion using a Continuous-Time Independent Cascade (CTIC) model~\cite{gomez2011uncovering}, in which diffusion is governed by misinformation contagiousness and users' susceptibility.
Within this model, we uniformly formalize three types of interventions---nudges, prebunking, and contextualization---as reductions in users' susceptibility, and calibrate the model using empirical diffusion and survey data.
This simulation-based framework allows us to systematically vary intervention strength, scale, timing, and targeting strategy, and to assess their collective-level implications under controlled diffusion dynamics.
In addition, we complement the simulations with theoretical analysis based on the Quenched Mean-Field (QMF) approximation~\cite{gomez2010discrete,prakash2012threshold,pastor2015epidemic} to aid the interpretation and generalization of the results.

\begin{table*}[t]
\centering
\small
\begin{tabular}{clcccccc}
\hline
Study 
& Diffusion model 
& I
& II
& III
& IV \\
\hline

\cite{bak2022combining} 
& Branching process-inspired generative model (no explicit network)
& $\times$
& $\checkmark$ 
& $\times$
& $\checkmark$ \\

\cite{zehmakan2023rumors} 
& IC-based rumor spreading model with trust similarity and forgetting
& $\checkmark$
& $\times$ 
& $\times$ 
& $\times$ \\

\cite{pilditch2022psychological} 
&  Agent-based belief diffusion model with Bayesian updating
& $\times$ 
& $\times$ 
& $\times$ 
& $\checkmark$ \\

\cite{Gausen2021can} 
& SIR-like compartmental rumor spreading model 
& $\times$ 
& $\checkmark$
& $\times$ 
& $\times$ \\

\cite{li2024large,liu2024skepticism,liu2025mosaic,liu2025stepwise,qiao2025dynamic,chuang2024simulating}
& LLM agent-based diffusion
& $\times$ 
& $\times$ 
& $\times$ 
& $\times$ \\

\textbf{Ours} 
& CTIC model
& $\checkmark$
& $\checkmark$
& $\checkmark$ 
& $\checkmark$ \\

\hline
\end{tabular}
\caption{
Comparison of existing studies and this work in terms of
(I) use of real-world network data,
(II) empirical calibration of diffusion dynamics,
(III) empirical calibration of intervention effects, and
(IV) systematic exploration of intervention parameters enabling counterfactual analysis.
}
\label{tab:related_work_comparison}
\end{table*}

Simulation results reveal that, under empirically estimated intervention levels, user interventions do not necessarily produce sufficient collective effects on networks.
For each intervention, misinformation prevalence monotonically decreases as intervention strength increases.
However, higher contagiousness substantially raises the intervention strength required to achieve substantial suppression.
In particular, single interventions have little impact on prevalence under the estimated contagiousness.
We also find that design adjustments such as increasing intervention scale, implementing interventions earlier, or strategically targeting high out-degree nodes are effective only when intervention strength is sufficiently large, while their effects remain limited at current intervention levels.
Furthermore, although combining multiple interventions increases the suppression effect, the reduction in misinformation prevalence remains bounded.
These findings quantitatively clarify the gap between micro-level behavioral change and macro-level diffusion suppression, and demonstrate the limitations of evaluating user interventions solely based on individual-level effects.

In summary, this paper makes the following contributions:
\begin{itemize}
\item We introduce an empirically calibrated simulation framework that bridges individual-level intervention effects and collective-level misinformation diffusion on real-world networks.
\item We quantitatively demonstrate that empirically observed individual-level intervention effects translate into only bounded reductions in network-level misinformation prevalence under realistic diffusion dynamics.
\end{itemize}

\section{Related Work} 

Several studies have investigated how micro-level user interventions influence the macro-level spread of misinformation using simulation-based approaches, although each addresses only a subset of relevant dimensions (Table~\ref{tab:related_work_comparison}).

Bak-Coleman et al.~\cite{bak2022combining} model the temporal dynamics of misinformation posts using large-scale Twitter data from the 2020 U.S. presidential election and evaluate interventions such as content removal, account banning, and nudges, both individually and in combination. 
This study is notable for calibrating diffusion dynamics using real-world data. 
However, intervention effects are introduced as exogenous manipulations rather than being estimated from empirical studies, and explicit network structure and intervention target selection are not considered.

Zehmakan et al.~\cite{zehmakan2023rumors} construct an independent cascade (IC)-based rumor diffusion model incorporating trust similarity and forgetting effects, and theoretically derive network conditions under which rumors spread rapidly while comparatively evaluating multiple intervention strategies. 
While this work explicitly models network diffusion, neither diffusion dynamics nor intervention effects are empirically calibrated.

Focusing on psychological interventions, Pilditch et al.~\cite{pilditch2022psychological} employ an agent-based model with Bayesian belief updating to evaluate how inoculation affects belief distributions across a population. 
Their framework enables systematic exploration of intervention parameters, but relies on synthetic networks and does not calibrate diffusion or intervention effects using real-world data.

Similarly, Gausen et al.~\cite{Gausen2021can} use Susceptible-Infected-Recovered (SIR)-like epidemiological compartmental model incorporating psychological inoculation and accuracy flags to compare intervention effects on misinformation spread. 
Although diffusion parameters are partially informed by empirical studies, intervention effects are not empirically calibrated and evaluations are conducted under limited parameter settings.

In recent years, LLM-based multi-agent simulations (LLMMAS) have been proposed to qualitatively study misinformation spread and its countermeasures by simulating natural language interactions among LLM agents with human-like text generation and reasoning capabilities~\cite{li2024large,liu2024skepticism,liu2025mosaic,liu2025stepwise,qiao2025dynamic,chuang2024simulating}. 
While these approaches enable rich behavioral representations, they currently face substantial challenges as frameworks for quantitatively evaluating macro-level diffusion dynamics, including limited verifiability, difficulties in empirical calibration, reproducibility concerns, and high computational cost~\cite{larooij2025large}. 
In particular, LLM agents lack explicit parameters corresponding to diffusion processes, making systematic comparison of intervention effects on large-scale networks difficult.

Taken together, existing studies provide important but partial insights into the collective effects of user-level misinformation interventions. 
In contrast, this study integrates 
(i) real-world network data,
(ii) empirically calibrated diffusion dynamics, 
(iii) empirically estimated intervention effects, and 
(iv) systematic exploration of intervention parameters within a unified diffusion framework. 
This enables comprehensive and counterfactual evaluation of how intervention strength, scale, timing, and targeting strategies shape macro-level misinformation prevalence.

\section{Misinformation Diffusion Model with User Interventions}
\label{sec:model}
\subsection{Modeling Misinformation Diffusion}
\label{sec:diffusion_model}

In this study, we employ the CTIC model~\cite{gomez2011uncovering} to describe the diffusion dynamics of misinformation on social networks.
The CTIC model is a continuous-time extension of the classical Independent Cascade model and assumes that transmission events along edges occur independently and probabilistically.
While more complex models that incorporate mechanisms such as social reinforcement, memory and forgetting, or multiple states can capture richer behavioral dynamics, they often entail difficulties in terms of calibration and interpretability.
Because our objective is to realistically and systematically evaluate the collective effects of interventions through empirically calibrated diffusion modeling, we adopt the CTIC model from the perspective of balancing realism and calibration feasibility.

We consider misinformation diffusion on a directed graph $G = (V, E)$, where $V$ is the set of nodes and $E$ is the set of edges.
In the CTIC model, each node is in one of two states: 
an \textit{inactive} state, in which the node has not yet received the misinformation, or an \textit{active} state, in which the node has received the misinformation and attempts to propagate it to other nodes.
When a node $u$ becomes active at time $t_u$, it attempts to propagate the misinformation exactly once, independently, to each inactive neighbor $v \in \Nout_u$. 
Each propagation attempt succeeds with probability $p_{uv}$, and if successful, node $v$ becomes active after a delay time $\tau_{uv}$. 
Once a node becomes active, its state does not change thereafter.

In this study, the propagation probability $p_{uv}$ on each edge $(u, v) \in E$ is defined as $p_{uv} = \eta \cdot s_v$, where $\eta \in [0,1]$ represents the contagiousness of the misinformation itself, and $s_v \in [0,1]$ represents the susceptibility of the receiving node $v$ to misinformation. 
Thus, $p_{uv}$ does not directly depend on the sender node $u$.
Indeed, for judging the credibility of fake content, the influence of the intermediary, that is, ``who shared it'', is limited compared to the source of the content or the characteristics of the receiver~\cite{shen2019fake}.
The propagation delay $\tau_{uv}$ on each edge $(u, v) \in E$ is assumed to follow an exponential distribution $\mathrm{Exp}(\lambda)$ with parameter $\lambda > 0$.

\subsection{Modeling User Interventions}
\label{sec:intervention_model}

In this study, to analyze how individual user interventions influence the final misinformation diffusion across the entire network, we incorporate three types of user interventions into the CTIC model: nudging, prebunking, and contextualization.

\subsubsection{Nudging}

Nudging is an intervention that aims to naturally steer users' beliefs and behaviors in a desirable direction by providing small psychological or behavioral incentives in users' decision-making processes, without prohibiting or mandating their actions~\cite{thaler2009nudge}. 
Representative nudges in misinformation countermeasures include accuracy prompts~\cite{pennycook2021shifting,pennycook2020fighting} and friction nudges~\cite{fazio2020pausing}. 
Accuracy prompts aim to suppress misinformation sharing behavior by directing users' attention to accuracy, for example by asking users to evaluate the accuracy of headlines or by presenting videos that emphasize the importance of sharing only accurate information. 
In contrast, friction nudges are interventions that suppress impulsive sharing behavior by introducing small amounts of effort or time costs into the act of sharing itself, such as displaying a confirmation screen that encourages reflection before sharing an article. 
Nudging is a relatively low-cost intervention that can be deployed at scale, and has been empirically shown to consistently and significantly reduce individual misinformation sharing behavior across demographic groups and topics~\cite{pennycook2020fighting,pennycook2022accuracy,kozyreva2024toolbox}.

Based on the above, in our model, nudging is defined as an operation that reduces the susceptibility of all users by a fraction $\varepsilon_{\nud}$ before the start of diffusion ($t = 0$), i.e., $s_v \to (1 - \varepsilon_{\nud}) s_v$. 
Note that although nudges in practice typically operate immediately before or during users' engagement with misinformation, rather than strictly before diffusion begins, applying them before the start of diffusion does not change the model behavior.

\subsubsection{Prebunking}

Prebunking is an intervention based on psychological inoculation theory~\cite{mcguire1961relative}, which aims to enhance users' cognitive immunity to misinformation by exposing users to refutations of imminent misinformation or teaching them typical manipulation techniques before they encounter the actual misinformation~\cite{van2020inoculating}.
Prebunking can be broadly categorised into passive inoculation, which provides users with information through text, infographics, and videos, and active inoculation, which enables users to understand misinformation manipulation techniques through proactive experiences such as gameplay.
Notably, active inoculation has been shown to be more effective than passive inoculation in suppressing the sharing of misinformation~\cite{basol2021towards}.
However, because prebunking requires voluntary user participation and involves a certain cognitive load, it is difficult to apply uniformly to all users, and its scalability is limited to small- to medium-scale deployments~\cite{kozyreva2024toolbox}.

Based on these empirical findings, in our model, prebunking is defined as an operation that reduces the susceptibility of a fraction $\delta_{\pre}$ of users by a fraction $\varepsilon_{\pre}$ before the start of diffusion ($t = 0$).

\subsubsection{Contextualization}
\label{sec:contextualization}

Contextualization is an intervention that aims to influence users' judgments and subsequent sharing behavior by adding background information, related context, warnings, or factual clarifications to misleading content circulating on social media platforms.
This category includes warning labels, fact-check notices, and related annotation-based mechanisms. 
A prominent example is Community Notes on Twitter/X, which allows users to collaboratively provide contextual information to misleading posts.
While Community Notes have been shown to influence post hoc corrective behaviors, such as voluntary retractions by original posters~\cite{gao2025can}, recent studies have increasingly examined their role in shaping judgments and sharing behavior of downstream users.
In particular, experimental evidence suggests that Community Notes improve users' ability to identify misleading content~\cite{drolsbach2024community} and reduce engagement and sharing with noted posts~\cite{slaughter2025community}.

Accordingly, we position contextualization not as a debunking intervention that corrects the beliefs of users who have already shared misinformation, but as a preventive intervention that issues warnings to users who have not yet shared misinformation and suppresses subsequent sharing behavior. 
Specifically, in our model, contextualization is defined as an operation that reduces the susceptibility of all users who have not yet shared misinformation by a fraction $\varepsilon_{\ctx}$ at time $t = T~(>0)$.

In practice, since the diffusion speed may vary depending on the contagiousness parameter $\eta$, applying an intervention at a fixed absolute time $T$ would compromise comparability across different values of $\eta$. 
To address this issue, in the experiments we introduce a parameter $\phi_{\ctx} \in [0,1]$ representing the diffusion stage, and define the intervention time $T = T(\eta, \phi_{\ctx})$ as
\begin{align*}
T(\eta,\phi_{\ctx})
= \inf \left\{ t \ge 0 \mid \rho_t(\eta) \ge \phi_{\ctx} \cdot \rho(\eta) \right\},
\end{align*}
where $\rho_t(\eta)$ denotes the misinformation prevalence at time $t$ without intervention, and $\rho(\eta)$ denotes the final prevalence without intervention.
This formulation enables comparison of intervention effects at equivalent diffusion stages across different values of $\eta$.

In this model, we assume that intervention effects combine independently and multiplicatively.
Accordingly, if a node $v$ receives all three interventions, its susceptibility is given by $(1 - \varepsilon_{\nud})(1 - \varepsilon_{\pre}) s_v$ at time $t < T$, and by $(1 - \varepsilon_{\nud})(1 - \varepsilon_{\pre})(1 - \varepsilon_{\ctx}) s_v$ at time $t \ge T$.
Although joint effects may deviate from this assumption in practice due to saturation or synergy, we adopt the multiplicative form as a simple baseline that enables systematic comparison and calibration.
We discuss this limitation in Section~\ref{sec:limitations}.

\subsection{Critical Condition from QMF Approximation}
\label{sec:qmf}

To theoretically understand how user interventions influence the critical condition of misinformation diffusion, we conduct an analysis based on the QMF approximation.
The QMF approximation is a mean-field approach that keeps the network structure fixed (quenched) while neglecting correlations between node states and treating influences between nodes independently.
It has been widely used to analyze critical conditions in epidemiological diffusion models~\cite{pastor2015epidemic}.

In the CTIC model with propagation delays $\tau_{uv} \sim \mathrm{Exp}(\lambda)$, the probability that node $v$ is active by time $t$, denoted by $\theta_v(t)$, can be expressed under the QMF approximation as
\begin{align}
    \theta_v(t) \approx \sum_{u \in \Nin_v} p_{uv} \int_0^t \lambda e^{-\lambda (t-s)} \theta_u(s)\,\dd s .
\end{align}
Assuming exponential growth $\theta_v(t) \sim e^{\gamma t}$ in the early stage of diffusion, we obtain the following critical condition:
\begin{align}
\Lambda_{\max}(\eta \bm{A}\,\diag(\bm{s})) = 1,
\end{align}
where $\bm{A} = [A_{vu}]$ denotes the adjacency matrix, $\diag(\bm{s})$ denotes the diagonal matrix of node susceptibilities, and $\Lambda_{\max}(\bm{M})$ denotes the largest eigenvalue of the matrix $\bm{M}$. 
Details of the derivation of the critical condition are provided in Appendix~\ref{sec:appendix}.

Under the QMF approximation, we can analytically evaluate the conditions on intervention parameters required for complete suppression of misinformation (i.e., $\rho \approx 0$), as well as differences in efficiency across intervention strategies.
In our model, both nudging and prebunking are defined as operations that reduce the susceptibility $s_v$ of all (or some) nodes to $s_v' = (1 - \varepsilon_\ast) s_v$ before the start of diffusion. 
Thus, the critical condition under nudging or prebunking intervention can be expressed as $\Lambda_{\max}(\eta \bm{A}\,\diag(\bm{s}')) = 1$, where $\diag(\bm{s}')$ is the post-intervention susceptibility matrix.
In contrast, contextualization is applied during the diffusion process and dynamically depends on the activation state of nodes.
Thus, it cannot be directly handled within the QMF approximation framework.

\section{Evaluating Collective Effects of User Interventions on Misinformation Diffusion}

In this section, we calibrate the diffusion parameters of the CTIC model and the intervention effects based on real-world Twitter diffusion data and publicly available user survey data on each intervention. 
Using the calibrated diffusion model, we then conduct numerical simulations to quantitatively evaluate how factors such as intervention strength, scale, timing, targeting strategy, and the combination of multiple interventions influence misinformation prevalence at the network level.
The experimental code and data are available at \url{https://github.com/s-furutani/macro_intervention_effect}.

\subsection{Simulation Setup}
\label{sec:setup}

\subsubsection{Network Data}

In this experiment, we use a real-world follower network data extracted from Twitter by Nikolov et al.~\cite{nikolov2020right}.
This network is constructed from a $10\%$ random sample of public tweets during the period from June 1 to 30, 2017.
The data includes users who shared at least 10 links to news sources, with at least one link being to a source labeled as low-credible.
For most accounts in the dataset, a misinformation score is defined as the proportion of posts containing links to low-credible sources.
In this study, we use this score as the susceptibility $s_v$ for each node.
We constructed a subgraph of the dataset consisting of $15{,}056$ nodes with misinformation scores and further extracted the largest weakly connected component.
As a result, the final network contains $14{,}991$ nodes and $4{,}327{,}446$ edges.

\subsubsection{Calibration of Diffusion Parameters}

Although the Nikolov dataset contains information on network structure and node susceptibility, it does not include diffusion histories of tweets and therefore cannot be used to estimate the parameters ($\eta$, $\lambda$) of the CTIC model. 
Accordingly, we instead use the Twitter diffusion dataset~\cite{hodas2014simple} for parameter estimation of the CTIC model.
This dataset is constructed from approximately three million tweets containing URLs collected in October 2010.
It includes timestamps for the sharing of $3{,}461$ tweets and a follower network composed of the original tweet authors and their followers ($12{,}627$ nodes and $619{,}262$ edges).
For parameter estimation, we use only $288$ relatively large cascades whose cascade sizes reached at least $100$ within $100$ hours.

\begin{figure}[t]
    \centering
    \includegraphics[width=1\linewidth]{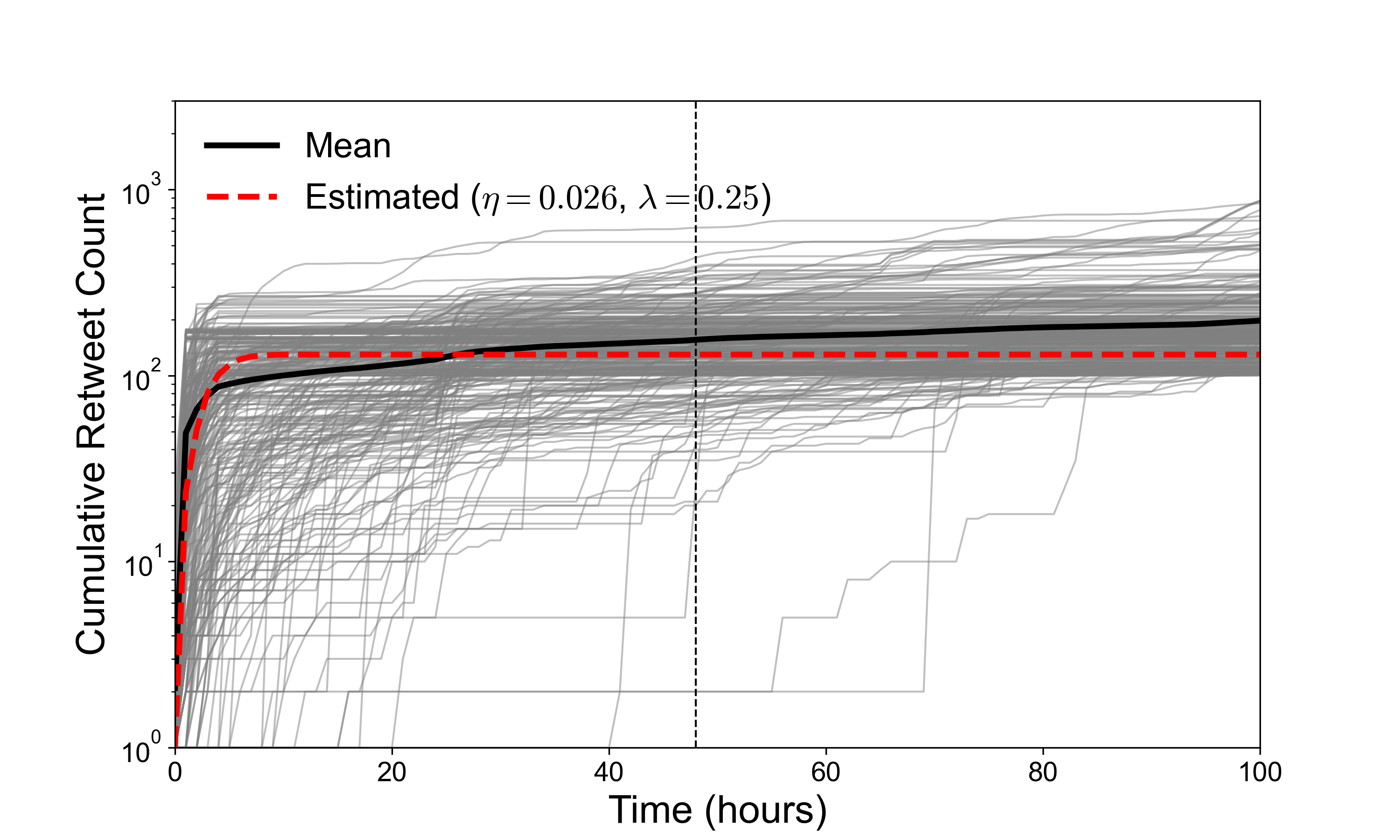}
    \caption{Empirical and simulated cumulative retweet counts over time.
    Gray lines represent individual empirical cascades, the solid black line represents their mean, and the dashed red line represents to the CTIC simulation with the estimated parameters. 
    Vertical dashed line indicates $t=48$.}
    \label{fig:twitter_fit}
\end{figure}

For parameter estimation, we first randomly assigned susceptibility $s_v$ to each node in this follower network such that its distribution matched the susceptibility distribution in the Nikolov dataset.
Next, for each cascade, we selected the user with the largest out-degree as the seed node and conducted information diffusion simulation using the CTIC model and recorded the cumulative retweet count over time $t$ since the start of diffusion.
To reproduce the average diffusion size and speed during the early growth phase, we estimated the contagiousness $\eta$ and delay rate $\lambda$ via grid search, minimizing the Euclidean distance loss between the simulation curve and the average cumulative retweet count based on empirical data for $t \in [0, 48]$.
As a result, we obtained estimates $\hat{\eta}=0.026$ and $\hat{\lambda}=0.25$.
Figure~\ref{fig:twitter_fit} shows the empirical cumulative retweet counts of each tweet and the simulation curve using the estimated parameters.

\subsubsection{Calibration of Intervention Effects}

In this study, to calibrate the strength of intervention effects based on empirical data, we use publicly available data from existing user survey studies on nudging, prebunking, and contextualization that satisfy the following conditions:
(i) a sufficiently large sample size ($N > 500$);
(ii) random assignment of participants to a control condition $C_0$ and an intervention condition $C_1$; and
(iii) availability of individual-level data on each participant $u$'s sharing willingness $z(a, u)$ toward misinformation content $a$.

Specifically, for nudging interventions, we use experimental data on accuracy prompts from Study~3 in~\cite{pennycook2021shifting} and on friction nudges in~\cite{fazio2020pausing}. 
In both studies, sharing willingness toward fake news headlines is measured on a 1--6 Likert scale.
For prebunking interventions, we use experimental data on active inoculation (\textit{GoViral}) and passive inoculation (\textit{Infographics}) from Study~2 in~\cite{basol2021towards}. 
In that study, sharing willingness toward COVID-19-related misinformation posts is measured on a 1--7 Likert scale.
For contextualization, we use experimental data on Community Notes in~\cite{drolsbach2024community}, and estimate intervention effects by comparing the \textit{No Flag} condition with the \textit{Community Note} condition. 
In this experiment, sharing willingness is measured as a binary outcome.
Note that all survey data are anonymized and do not contain personally identifiable information.

In our model, the strength of an intervention can be expressed as the relative reduction in user susceptibility induced by the intervention: 
\begin{align}
    \varepsilon_{\ast} = \frac{s_v - s_v'}{s_v}. 
\end{align}
Thus, we estimate intervention strength by treating reported sharing willingness as a proxy for susceptibility and computing the relative reduction in average sharing willingness toward misinformation content between the control and intervention conditions.
First, because sharing willingness $z(a, u)$ of participant $u$ toward misinformation content $a$ is measured on different scales across datasets, we linearly rescale all values to the interval $[0, 1]$. 
For example, for a 1–6 Likert scale, we apply the transformation $(z - 1) / 5$.
Next, for each misinformation content item $a$, we compute the average sharing willingness under the control condition as $\bar{z}^0(a) = \sum_{u \in C_0} z(a, u) / N_0$ and under the intervention condition as $\bar{z}^1(a) = \sum_{u \in C_1} z(a, u) / N_1$, where $N_0$ and $N_1$ denote the numbers of participants assigned to the control and intervention conditions, respectively.
Using these quantities, we compute the suppression rate for misinformation content $a$ as
\begin{align}
    e(a) = \frac{\bar{z}^0(a) - \bar{z}^1(a)}{\bar{z}^0(a)} .
\end{align}
Since the ratio becomes unstable when $\bar{z}^0(a)$ is extremely small, we exclude items with $\bar{z}^0(a) < 0.10$ from the analysis. 

\begin{figure}[t]
    \centering
    \includegraphics[width=1\linewidth]{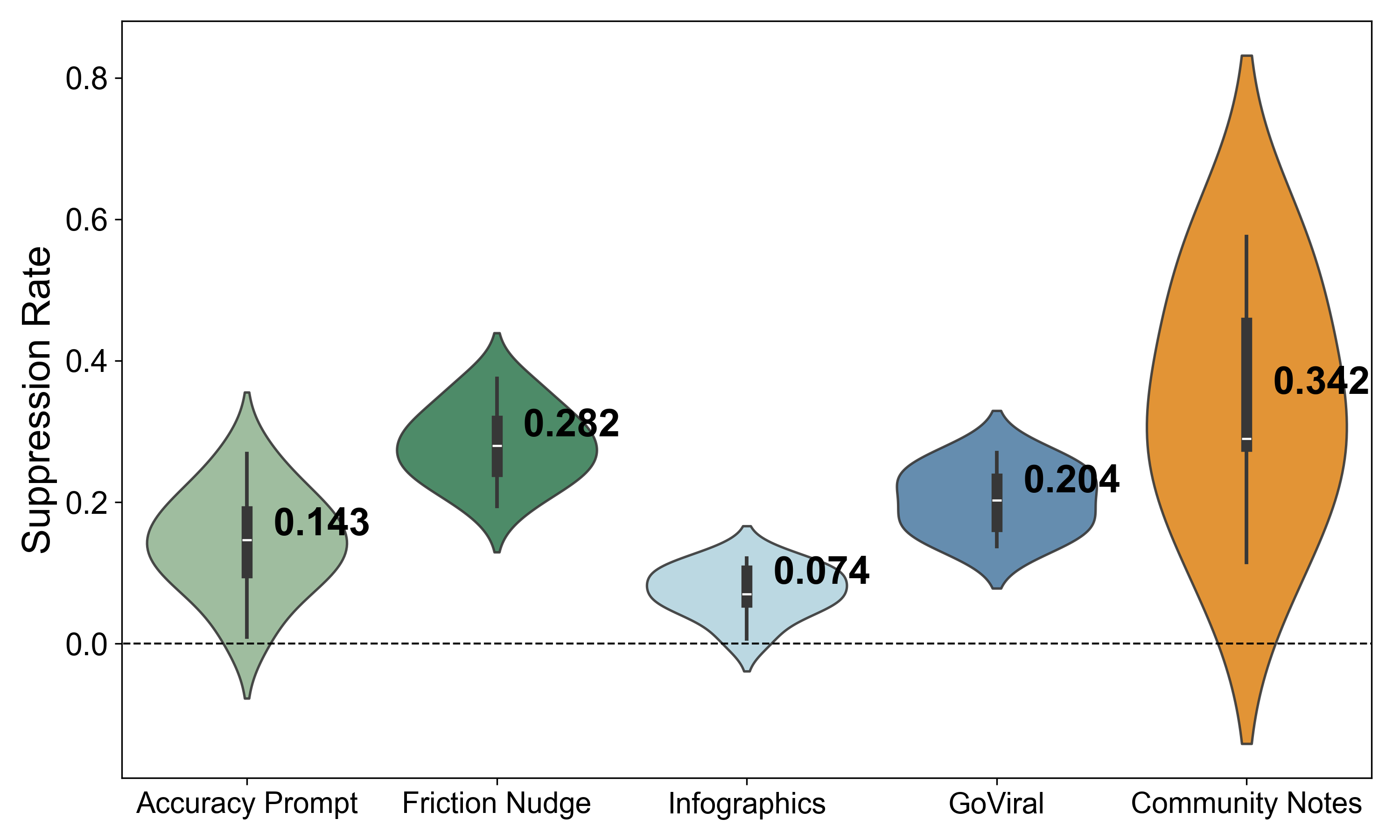}
    \caption{Distributions of content-level suppression rates $e(a)$ for each intervention. 
    Numbers indicate mean values.}
    \label{fig:epsilon_fit}
\end{figure}

Figure~\ref{fig:epsilon_fit} shows the distribution of suppression rates of each intervention, where the values reported in the figure correspond to the mean suppression rate of each intervention.
Based on these results, we set the estimated intervention strengths for nudging, prebunking, and contextualization to $\hat{\varepsilon}_{\nud} = 0.143$, $\hat{\varepsilon}_{\pre} = 0.204$, and $\hat{\varepsilon}_{\ctx} = 0.342$, respectively.

\subsubsection{Other Parameter Settings}

To reflect the reachability of interventions in the real world, we set the parameters related to the intervention scale of prebunking and the intervention timing of contextualization based on reported values from existing studies.

First, for the intervention scale, we refer to a large-scale field experiment on the YouTube platform using inoculation video advertisements~\cite{roozenbeek2022psychological}. 
The study reports that among approximately $5.4$ million YouTube users, the proportion of users who watched the inoculation video for 30 seconds or more was about $0.179~(\simeq 967{,}347 / 5{,}400{,}000)$. 
Thus, we adopt $\hat{\delta}_{\pre} = 0.2$ as a baseline value representing a realistic scale of prebunking interventions.

Next, regarding the timing of contextualization, prior work reports that the average half-life of engagement with posts on Twitter is approximately $79.5$ minutes~\cite{pfeffer2023half}, whereas Community Notes are displayed, on average, several days after a post is published~\cite{chuai2024did}. 
Accordingly, we assume a situation in which contextualization is applied after diffusion has substantially progressed, and set $\hat{\phi}_{\ctx} = 0.8$ as the baseline value.

In addition, in the experiments we consider a scenario in which an influential user actively spreads misinformation. 
In the simulations, we select as the seed node the node with susceptibility $s_v = 1$ (i.e., a node whose proportion of shared links to low-credibility sources in the dataset is $100\%$) that has the largest out-degree.

\subsection{Simulation Results}
\label{sec:results}

In this section, we first compare the misinformation prevalence as functions of intervention strength and contagiousness for each intervention type.
Next, for prebunking and contextualization interventions, we examine the effects of intervention scale and intervention timing. 
We then compare different target selection strategies for prebunking interventions, and finally present the effects of combining multiple interventions under the realistic conditions.

\begin{figure*}[t]
  \centering
  \includegraphics[width=0.9\linewidth]{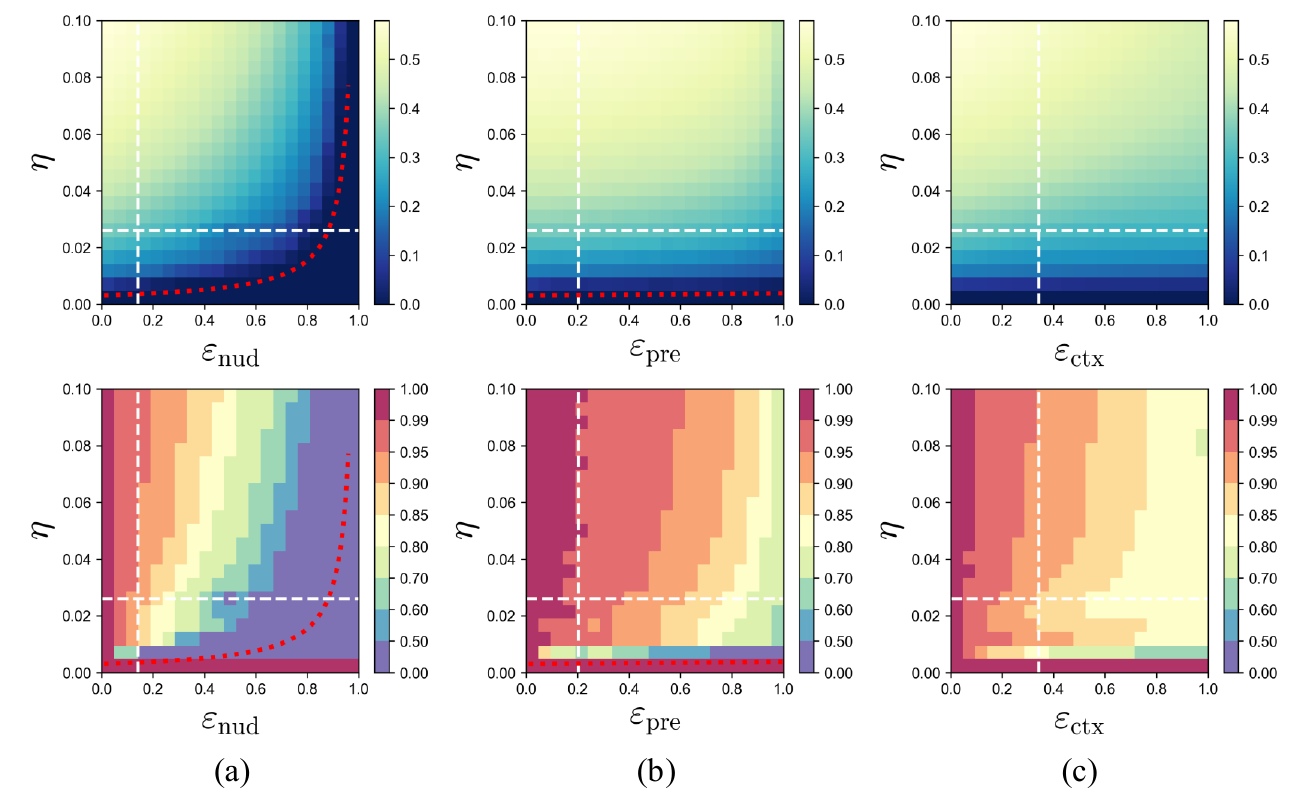}
  \caption{Misinformation prevalence (top row) and relative prevalence (bottom row) as functions of intervention strength and contagiousness for (a) nudging, (b) prebunking, and (c) contextualization interventions.
  White dashed lines indicate estimated parameter values.
  Red dotted lines indicate the critical curves predicted by the QMF approximation.}
  \label{fig:macro_effect_vary_eta}
\end{figure*}

\subsubsection{Effects of Intervention Strength and Contagiousness}

Figure~\ref{fig:macro_effect_vary_eta} shows the misinformation prevalence $\rho(\varepsilon_{\ast}, \eta)$ and the relative prevalence $\tilde{\rho}(\varepsilon_{\ast}, \eta)$ on the Nikolov network.
Here, the relative prevalence is defined as the ratio of the misinformation prevalence to that in the absence of intervention ($\varepsilon_{\ast} = 0$), i.e., $\tilde{\rho}(\varepsilon_{\ast}, \eta) = \rho(\varepsilon_{\ast}, \eta) / \rho(0, \eta)$.
We set $\delta_{\pre} = \hat{\delta}_{\pre}$ and $\phi_{\ctx} = \hat{\phi}_{\ctx}$ as given in Section~\ref{sec:setup}.

First, we find that all interventions have a modest effect on reducing the prevalence of misinformation under the estimated contagiousness and intervention strengths from real-world dataset.
For all intervention types, misinformation prevalence decreases monotonically as the intervention strength $\varepsilon_{\ast}$ increases, whereas higher contagiousness $\eta$ requires stronger interventions to achieve the same level of suppression.
This trend is especially pronounced in the critical curves shown in the figure. 
For nudging, the lower bound on the intervention strength required to achieve complete suppression of misinformation ($\rho(\varepsilon_{\ast}, \eta) \approx 0$) increases sharply as contagiousness increases. 
For prebunking with an intervention scale of $\delta_{\pre} = \hat{\delta}_{\pre}$, complete suppression is difficult except when contagiousness is extremely low.

\subsubsection{Effects of Intervention Scale and Timing}

\begin{figure}[t]
  \centering
  \includegraphics[width=1\linewidth]{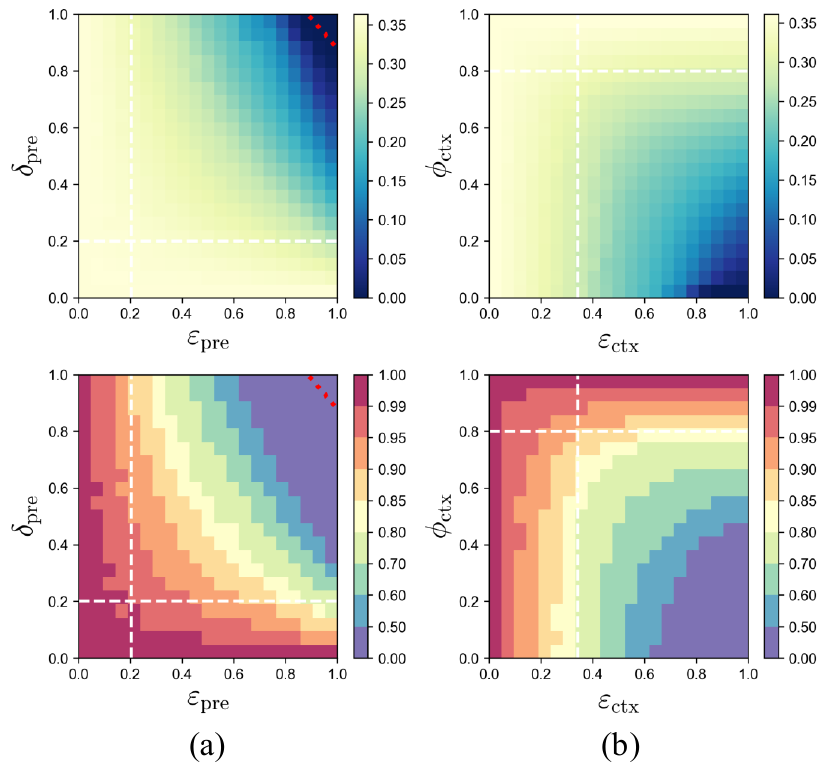}
  \caption{Misinformation prevalence (top row) and relative prevalence (bottom row) as functions of intervention strength and intervention scale for prebunking (a), and intervention strength and intervention timing for contextualization (b).
  White dashed lines indicate estimated parameter values.
  Red dotted lines indicate critical curves predicted by the QMF approximation.
}

  \label{fig:macro_effect_fixed_eta}
\end{figure}

Figure~\ref{fig:macro_effect_fixed_eta} shows the misinformation prevalence and relative prevalence for prebunking and contextualization interventions for the intervention parameter pairs $(\varepsilon_{\pre}, \delta_{\pre})$ and $(\varepsilon_{\ctx}, \phi_{\ctx})$, respectively, with fixed $\eta = \hat{\eta}$.
These figures indicate that, for both prebunking and contextualization, the influence of intervention scale $\delta_{\pre}$ and intervention timing $\phi_{\ctx}$ depends on the magnitude of intervention strength $\varepsilon_{\ast}$. 
In the region where intervention strength $\varepsilon_{\ast}$ is small, moderate changes in $\delta_{\pre}$ or $\phi_{\ctx}$ have little effect on misinformation prevalence. 
In contrast, in the region where intervention strength $\varepsilon_{\ast}$ is large, changes in $\delta_{\pre}$ or $\phi_{\ctx}$ have a substantial influence on prevalence, with interventions applied to more users or at earlier stages leading to greater suppression of misinformation diffusion.
However, according to our estimates, current interventions do not have sufficiently large intervention strength, and therefore increasing the number of intervened users or advancing the timing of intervention is not expected to suppress misinformation prevalence.

\subsubsection{Effects of Intervention Strategies}

Next, we evaluate how different target selection strategies affect intervention effectiveness.
We consider three strategies:
a degree-based strategy that prioritizes nodes with high out-degree,
a susceptibility-based strategy that targets nodes with high susceptibility,
and a distance-based strategy that intervenes on nodes close to the seed, surrounding the source of misinformation.

\begin{figure*}[t]
  \centering
  \includegraphics[width=0.9\linewidth]{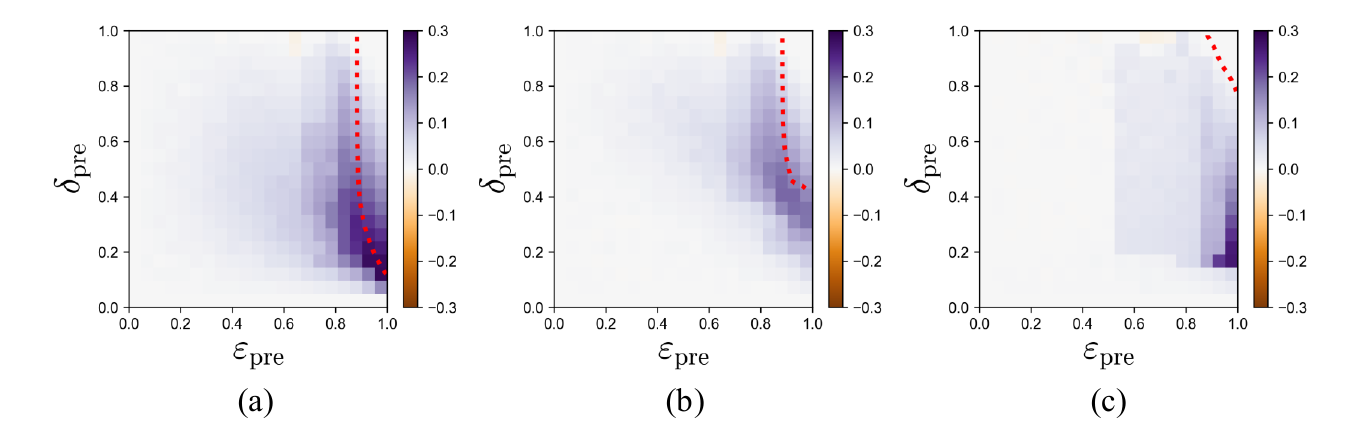}
  \caption{Differences in misinformation prevalence relative to random target selection, $\Delta \rho_{\mathrm{X}}(\varepsilon_{\pre}, \delta_{\pre})$, for prebunking interventions with (a) degree-based, (b) susceptibility-based, and (c) distance-based target selection strategies. 
  Red dotted lines indicate the critical curves predicted by the QMF approximation.}

  \label{fig:target_selection_diff}
\end{figure*}

\begin{figure*}[t]
    \centering
    \includegraphics[width=1\linewidth]{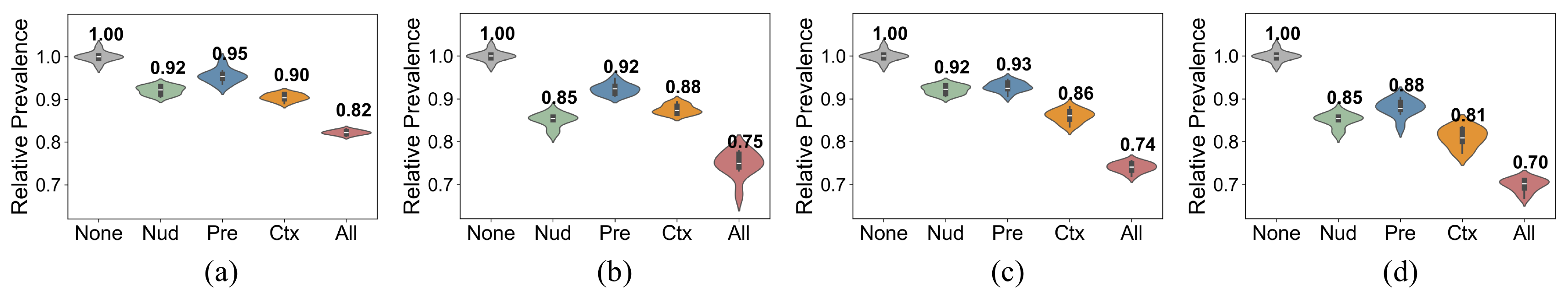}
    \caption{Relative misinformation prevalence under single interventions and their combination across four settings:
    (a) baseline empirical intervention parameters,
    (b) increased intervention strength,
    (c) expanded intervention reach, and
    (d) simultaneous improvements in both strength and reach.
    Numbers indicate mean values.
}
    \label{fig:single_vs_combined}
\end{figure*}

Figure~\ref{fig:target_selection_diff} shows the difference in misinformation prevalence relative to random selection (Fig.~\ref{fig:macro_effect_fixed_eta}a), defined as $\Delta \rho_{\mathrm{X}}(\varepsilon_{\pre}, \delta_{\pre}) = \rho_{\mathrm{Rand}}(\varepsilon_{\pre}, \delta_{\pre}) - \rho_{\mathrm{X}}(\varepsilon_{\pre}, \delta_{\pre})$ for each target selection strategy $\mathrm{X}$.
Overall, when $\varepsilon_{\pre}$ is small, the difference from random selection is negligible, and the effect of target selection strategies does not emerge. 
As $\varepsilon_{\pre}$ increases, however, the influence of target selection gradually becomes apparent. 

Compared to other strategies, the degree-based strategy most effectively suppresses misinformation prevalence.
In the CTIC model, nodes with high out-degree generate more secondary infections on average, and thus contribute disproportionately to diffusion.
Thus, prioritizing such nodes therefore yields the largest reduction in network-level prevalence.
The susceptibility-based strategy achieves the second-largest suppression effect, as reducing susceptibility of highly susceptible nodes directly lowers the average transmission probability.

Notably, a distance-based strategy similar to ring immunization, which is commonly used in infectious disease control, proves remarkably ineffective in this setting.
This is because, unlike the biological vaccination that provides high infection prevention efficacy, the prebunking intervention only slightly reduces misinformation susceptibility.
Thus, even if the source of misinformation is surrounded by prebunked users, it remains difficult to prevent its outward spread.
These results suggest that, for suppressing the spread of misinformation, simply importing analogies from infectious disease control may not work, and careful design considering the differences between the two is necessary.

Consistent with the above results, the critical curves indicate that complete suppression is achievable with the smallest intervention scale under the degree-based strategy, followed by the susceptibility-based and distance-based strategies.

\subsubsection{Effects of Combined Interventions}

We here evaluate the effects of combining multiple interventions under the following four intervention parameter settings:
(i) the current, empirically estimated intervention parameters specified in Section~\ref{sec:setup} (i.e., $\hat{\eta} = 0.026$, $\hat{\lambda} = 0.25$, $\hat{\varepsilon}_{\nud} = 0.143$, $\hat{\varepsilon}_{\pre} = 0.204$, $\hat{\varepsilon}_{\ctx} = 0.342$, $\hat{\delta}_{\pre} = 0.2$, and $\hat{\phi}_{\ctx} = 0.8$);
(ii) a modest increase in intervention strength (i.e., $\varepsilon_\ast = \hat{\varepsilon}_\ast + 0.1$);
(iii) a modest increase in intervention reach through expanded scale and earlier deployment (i.e., $\delta_\pre = \hat{\delta}_\pre + 0.1$ and $\phi_\ctx = \hat{\phi}_\ctx - 0.1$); and
(iv) simultaneous improvements in both intervention strength and reach.

Figure~\ref{fig:single_vs_combined} shows the relative misinformation prevalence when each intervention is applied individually and when multiple interventions are combined under these four settings.
Under the current intervention levels (Fig.~\ref{fig:single_vs_combined}a), each individual intervention yields only a modest suppression effect, reducing misinformation prevalence by approximately $5$--$10\%$ compared to the no-intervention baseline.
In contrast, when all interventions are combined, misinformation prevalence is reduced by approximately $18\%$.
When intervention strength is increased (Fig.~\ref{fig:single_vs_combined}b), or when intervention reach is improved (Fig.~\ref{fig:single_vs_combined}c), misinformation prevalence can be reduced by approximately $25$--$26\%$ with combined interventions.
When both intervention strength and reach are simultaneously improved (Fig.~\ref{fig:single_vs_combined}d), the overall misinformation prevalence is reduced by approximately $30\%$.

Taken together, these results indicate that combining multiple interventions is more effective at suppressing misinformation than applying interventions individually, as already reported in~\cite{bak2022combining}.
However, under current intervention levels, substantial suppression remains difficult to achieve.
Moreover, even when intervention strength and reach are modestly improved within realistic bounds, the maximum reduction in misinformation prevalence remains limited to around $30\%$.
This suggests that substantially suppressing misinformation spread through user-level interventions alone is challenging, and that complementary system-level countermeasures beyond user interventions may be necessary.

\section{Discussion}

\subsection{Interpretation of Results and Implications}
\label{sec:implications}

Our results demonstrate that user interventions that are statistically significant at the individual level do not necessarily translate into substantial suppression of misinformation prevalence at the collective level.
In particular, at empirically estimated intervention levels, single interventions have little impact on prevalence under the estimated contagiousness.
Moreover, design adjustments such as increasing intervention scale, intervening earlier, or strategically targeting high-degree users become effective only when intervention strength is sufficiently large.
Combining multiple interventions is a promising direction and improves suppression relative to applying each intervention alone.
However, under current and realistically attainable intervention levels, the reduction in misinformation prevalence remains bounded at around $20$--$30\%$.

These results highlight a fundamental gap between individual-level and collective-level effects of misinformation interventions, suggesting that evaluating interventions solely at the individual level can lead to overly optimistic expectations regarding their societal impact.
More broadly, they also indicate that comparing interventions only by their individual-level effect sizes (i.e., how much they reduce susceptibility among treated users) can be misleading when the goal is to curb network-level prevalence.
The collective impact is shaped not only by intervention strength, but also by deployability, that is, how widely an intervention can be applied, how early it can be triggered during diffusion, and which users it can realistically target.
For instance, an intervention with a larger individual-level effect may yield a smaller collective benefit if it can be delivered to only a limited fraction of users, whereas a weaker intervention with broader reach can outperform it at the network level.
Similarly, contextualization mechanisms may exhibit strong preventive effects in user studies, yet their collective impact can be limited if they are typically activated only after diffusion has already progressed.
Overall, these results suggest that evaluation and design should focus on the collective-effect profile of an intervention, such as strength together with reach, timing, and targeting constraints, rather than effect size alone.

Importantly, these findings should not be interpreted as arguing against user-level interventions.
Rather, they suggest that there is no single ``silver bullet'' intervention capable of substantially reducing misinformation prevalence on its own.
Meaningful suppression is likely to require carefully coordinated combinations of interventions spanning both user-level and system-level measures.

\subsection{Limitations}
\label{sec:limitations}

\subsubsection{Methodological Scope and Trade-offs}
This study adopts a simulation-based approach to evaluate the collective effects of user-level misinformation interventions.
Alternative approaches to the same research objective exist, including causal analyses of observational social media data~\cite{chuai2024did,chuai2024community,hwang2025nudge} and platform-scale field experiments~\cite{pennycook2021shifting,roozenbeek2022psychological,katsaros2022reconsidering,kurek2025follow}.
While these approaches provide valuable real-world evidence, they often have limitations in their ability to systematically vary intervention parameters or conduct counterfactual analysis under controlled conditions.
Numerical simulation, while involving simplifying assumptions, provides a complementary perspective by enabling systematic exploration of intervention design parameters based on empirically calibrated diffusion dynamics.

\subsubsection{Modeling Assumptions and Data Limitations}
First, we rely on the CTIC model to represent misinformation diffusion. 
As a result, the model abstracts away several real-world mechanisms, including social reinforcement, emotional responses, evolving user attention, cross-platform information flows, offline exposure, and interactions among multiple competing pieces of content. 
While these simplifications may overlook some qualitative aspects of real-world misinformation dynamics, they enable transparent empirical calibration and systematic analysis of how intervention design parameters shape collective diffusion outcomes.

Second, the network and diffusion data used in this study are based on previously collected Twitter/X datasets. 
While such data remain widely used due to current API access restrictions, they may not fully reflect recent platform dynamics. 
However, our goal is not to predict absolute prevalence levels but to analyze relative effects and structural relationships between interventions and diffusion outcomes.

Third, intervention effects are calibrated using existing survey and experimental studies. 
These estimates may depend on population characteristics, content domains, and experimental settings, and therefore may not generalize uniformly across contexts.

Finally, when evaluating combinations of interventions, we assume that multiple interventions act independently and multiplicatively on users' susceptibility.
Under this assumption, stacking interventions monotonically reduces susceptibility and can, in principle, drive it arbitrarily close to zero.
In practice, however, intervention effects may exhibit saturation, such that additional interventions yield diminishing returns rather than unlimited reductions~\cite{fujimoto2025joint}.
Conversely, multiple interventions may also interact synergistically, exerting an impact greater than the multiplicative reduction assumed in our model~\cite{pennycook2024inoculation}.
As a result, our model may misestimate the effectiveness of combining multiple interventions.

\section{Conclusion}

In this study, we systematically evaluated how individual-level user interventions affect collective-level misinformation prevalence through numerical simulations using an empirically calibrated misinformation diffusion model. 
Our results show that although these interventions consistently reduce misinformation prevalence as their strength increases, higher contagiousness substantially raises the intervention strength required to achieve meaningful suppression. 
Under empirically estimated intervention levels, even systematically adjusted intervention designs yield only moderate collective reductions in misinformation prevalence.

The main contribution of this work lies in quantitatively demonstrating the gap between micro-level behavioral intervention effects and macro-level misinformation diffusion outcomes. 
By situating individual-level findings within a network diffusion framework, this study clarifies both the potential and the limitations of user interventions, highlighting the need to assess their effectiveness beyond individual-level outcomes.

\bibliographystyle{unsrt}
\bibliography{ref}

\appendix 
\section{Derivation Details of the Critical Condition}
\label{sec:appendix}

Let $\theta_v(t)$ denote the probability that node $v$ is active by time $t$. 
When the delay on each edge follows an exponential distribution $\mathrm{Exp}(\lambda)$, the probability that a node $u$ activated at time $s$ activates node $v$ by time $t$ is given by $p_{uv} (1 - e^{-\lambda (t - s)})$. 
In addition, the probability density that node $u$ becomes active in the infinitesimal interval $[s, s+\dd s]$ can be written as $\frac{\dd \theta_u(s)}{\dd s}$.
Under these assumptions, the probability $Q_{uv}(t)$ that no transmission occurs along edge $(u, v)$ by time $t$ is given by
\begin{align*}
    Q_{uv}(t) &= 1 - \int_0^t p_{uv} (1 - e^{-\lambda (t - s)}) \frac{\dd \theta_u(s)}{\dd s} \dd s \\
    &= 1 - p_{uv} \int_0^t \lambda e^{-\lambda (t - s)} \theta_u(s) \dd s ,
\end{align*}
where the second equality follows from integration by parts and the condition $\theta_u(0) = 0$.
Assuming that transmissions from different nodes occur independently, we obtain
\begin{align*}
    \theta_v(t) = 1 - \prod_{u \in \Nin_v} Q_{uv}(t)
    = 1 - \exp\!\left( \sum_{u \in \Nin_v} \ln Q_{uv}(t) \right) .
\end{align*}
Using the approximation $\ln(1 - x) \approx -x$ for $x \ll 1$, this can be written as
\begin{align*}
    \theta_v(t) \approx 1 - \exp\!\left(
    - \sum_{u \in \Nin_v} p_{uv} \int_0^t \lambda e^{-\lambda (t - s)} \theta_u(s) \dd s
    \right) .
\end{align*}
Applying a linear approximation yields
\begin{align*}
    \theta_v(t) \approx \sum_{u \in \Nin_v} p_{uv} \int_0^t \lambda e^{-\lambda (t - s)} \theta_u(s) \dd s .
\end{align*}
In the early stage of diffusion, assuming exponential growth $\theta_u(s) = \psi_u e^{\gamma s}$, we obtain, in the limit $t \to \infty$,
\begin{align*}
    \psi_v e^{\gamma t}
    &\approx \sum_{u \in \Nin_v} p_{uv} \int_0^t \lambda e^{-\lambda (t - s)} \psi_u e^{\gamma s} \dd s \\
    &= \frac{\lambda}{\lambda + \gamma} \sum_{u \in \Nin_v} p_{uv} \psi_u e^{\gamma t} \left( 1 - e^{-(\lambda + \gamma) t} \right) \\
    &\to \frac{\lambda}{\lambda + \gamma} \sum_{u \in \Nin_v} p_{uv} \psi_u e^{\gamma t} .
\end{align*}
Defining the transmission probability matrix as $\bm{P} = [p_{vu}]$, the above expression can be written as the eigenvalue equation
\begin{align*}
    \bm{P} \bm{\psi} = \frac{\lambda + \gamma}{\lambda} \, \bm{\psi} .
\end{align*}
Because the growth of the system is governed by the largest eigenvalue $\Lambda_{\max}(\bm{P})$, we obtain $\gamma = \lambda (\Lambda_{\max}(\bm{P}) - 1)$, which leads to the critical condition
\begin{align*}
    \Lambda_{\max}(\bm{P}) = \Lambda_{\max}(\eta \bm{A}\, \diag(\bm{s})) = 1 .
\end{align*}

\end{document}